# The MolecularWeb Universe: Web-Based, Immersive, Multiuser Molecular Graphics And Modeling, for Education and Work in Chemistry, Structural Biology, and Materials Sciences

*Luciano A. Abriata\**

Laboratory for Biomolecular Modeling and Protein Structure Core Facility, School of Life Sciences, École Polytechnique Fédérale de Lausanne (EPFL) and Swiss Institute of Bioinformatics (SIB), CH-1015 Lausanne, Switzerland

\* luciano.abriata@epfl.ch  https://lucianoabriata.altervista.org

**Abstract:**

For decades, software for molecular visualization has been a cornerstone of research and education in the chemical and structural sciences. It is unfortunate however that consumer devices enable only flat 2D inputs and outputs, which brings two problems. First, traditional 2D screen-mouse interfaces struggle to convey the inherently three-dimensional nature of molecules and their interactions. Second, probably worse, thoughtful manipulation of molecular structures in three dimensions is very hard to achieve with 2D peripherals. Immersive technologies under the umbrella of eXtended Reality (XR, including Augmented and Virtual Reality, AR and VR) promise to revolutionize how we learn, teach, and conduct research in the chemical and structural sciences, by assisting both how we see and how we manipulate molecules in 3D. This chapter chronicles the development of the "MolecularWeb" ecosystem, a suite of web-based tools designed to make immersive molecular visualization and interactive modeling and simulations widely accessible to a broad audience. Centralized at https://molecularweb.org, these tools have found wide applicability in education and science communication, with its next tools intended to directly assist research as well. We describe first moleculARweb, which delivers activities and tools for education in pass-through and mirror-like AR using regular devices like phones, tablets and computers. We then cover MolecularWebXR, a platform for multiuser immersion using WebXR-enabled headsets—yet remaining accessible with simpler devices—that brings content about chemistry, materials sciences and structural biology in formats engaging for educational and science outreach activities, also serving for immersive scientific discussions. A short visit to PDB2AR will show how users can create tailored content for MolecularWebXR and also stand-alone AR and VR experiences. We finally delve into our HandMol prototypes, which allow for immersive visualization of PDB files and their modeling with bare hands in immersive 3D by multiple concurrent users assisted by on-the-fly molecular mechanics calculations, seamless file exchange, a language model module that interprets users' inputs without them having to learn any menus or scripting, alternative immersive access with consumer devices, and more. With the various tools presented, we offer a glimpse into the present and near future of accessible, interactive and intuitive molecular science on the web.

## 1. Introduction: The challenge of visualizing and interacting with virtual representations of molecular systems

Understanding the three-dimensional structure, dynamics, and interactions of molecules is fundamental to a vast array of scientific disciplines, including chemistry, drug discovery, materials science, and molecular biology. For generations, scientists and science students have relied on physical models and two-dimensional (2D) computer graphics to visualize molecular systems and other entities of inherently 3D nature, such as molecular orbitals, surfaces, volumetric maps, etc.[1–4] While computer graphics have undeniably been invaluable tools, they present inherent limitations. On the one hand, there is the difficulty of representing and perceiving true depth and complex spatial relationships on a 2D display,[3,4] which traditional molecular graphics programs have tackled by allowing users to use cuing, clipping, cutting planes, etc. In fact, for users who gain dexterous use of such tools after months or years of work, the problem of realistic 3D perception on flat screens is probably overcome. However, this is not the case for novice users and for few-times users such as students or the public.

On the other hand, there is the difficulty in manipulating 3D objects with a 2D input peripheral such as a mouse, sometimes also with a keyboard, all of which is often cumbersome and counterintuitive even for experienced users. Achieving more natural manipulation of 3D objects in real 3D space can be particularly useful for intricate tasks such as molecular docking, detailed conformational analysis, or deciphering stereochemistry and chirality.[5–7] In fact we have argued in a recent perspective that the toughest limitation of traditional molecular graphics software is not in 3D visualization but in 3D manipulation.[8]

Immersive technologies, termed also XR after eXtended Reality and encompassing Virtual Reality (VR) which plunges the user into a completely digital environment, and Augmented Reality (AR) which overlays digital information onto the real world, offer a paradigm shift in human-computer interaction. Particularly for molecular graphics and modeling, XR-based applications promise a more natural and intuitive way to interact with 3D molecules and data associated with them.[6,7,9–12,12–21] The potential benefits for education are immense, offering students a more tangible, hands-on experience with abstract concepts that are traditionally difficult to grasp; for research, in turn, these technologies could accelerate discovery by allowing more fluid exploration and discussion of molecular interactions, properties, and dynamic behaviors.[6,11,22–30]

The literature on AR/VR systems for molecular graphics and XR in science is very extensive, with various recent reviews summarizing different parts of it.[14,23] Historically, however, the widespread adoption of immersive molecular modeling has been hampered by several factors. Firstly, high-end VR/AR systems were until recently prohibitively expensive for many end users, and some still are even in wealthy countries. Second, now largely alleviated, headsets used to entail bulky hardware tethered to external computers and in some cases also to towers equipped with instrumentation for body tracking. Third, the setup and use of these systems frequently involved, and in many cases still today involves, software configurations that represent a very steep barrier for adoption. Fourth, many early or simpler AR solutions were restricted to static visualizations, lacking the dynamic interactivity crucial for deeper understanding and research. Finally, a fragmented landscape of diverse hardware and software ecosystems made it difficult to develop universally accessible and standardized tools; in fact even today, most devices have different operating systems that require specialized compilations for software to run on them—think Apple vs. Android systems for smartphones and tablets, Windows vs. Mac vs. Linux for computers, or the bunch of different operating systems for the different AR/VR headsets. To the rescue and as we advocate for,[31–34] web programming and in particular the WebXR standard offer out-of-the-box instant access to immersive molecular graphics and modeling without installs and with supreme device compatibility.[31,35] All our tools from the MolecularWeb universe benefit from this technology, as we will cover in more detail throughout this chapter.

In what follows we present the MolecularWeb ecosystem in a chronological order that shows how we progressively enhanced, with a fully web-based spirit, visual immersion, seamless interactivity, and human-human plus human-computer collaboration applied to the molecular sciences. Our resources and platforms are all delivered through the web browser thus being available in every computer, phone, tablet, and the last generations of XR headsets from major vendors. Centralized at https://www.molecularweb.org, they include moleculARweb, the MolecularWebXR platform, the PDB2AR tool, the HandMol client-only prototype, and the upcoming MolecularWebXR-integrated HandMol (Table 1).

**Table 1. Web apps and web platforms of the molecularweb universe**

| Website/tool (section in this chapter) | Purpose/Use cases | Direct link(s) | References |
|---|---|---|---|
| **moleculARweb** (Section 2) | • Passthrough and mirror AR for webcam-enabled devices (laptops, smartphones, tablets) using printed fiducial markers (or no markers in some activities).<br>• Education in general and organic chemistry; introductory materials sciences and structural biology. | https://molecularweb.epfl.ch<br><br>https://molecularweb.org/ | 27,36,37 |
| **PDB2AR** (Section 3) | • Content generator for molecular visualization in AR with or without fiducial markers, in stand-alone VR on headsets, or inside MolecularWebXR rooms. | https://molecularweb.epfl.ch/pages/pdb2ar.html | 38 |
| **MolecularWebXR** (Section 4) | • Multiuser discussions in rooms with preset content or with content customized via PDB2AR. Virtual models are static only.<br>• Preset content targets introductory chemistry and structural biology. | https://molecularwebxr.org | 39 |
| **Client-only HandMol prototype** (Section 5 intro) | • Immersive visualization of PDB files by 1 or 2 concurrent users. Hand-driven molecule manipulation. On-request minimization with molecular mechanics.<br>• Immersive substitute of molecular graphics programs; suitable for small operations, education, etc. | https://go.epfl.ch/handmol | 40 |
| **MolecularWebXR-integrated HandMol prototype** (Section 5) | • Like MolecularWebXR but with molecules loaded from PDB files and parametrized for interactive simulations.<br>• Immersive substitute of molecular graphics software, enhanced with molecular mechanics, tools for collaboration, AI assistant for seamless control.<br>• For self-education, single-person work, teacher-student(s) sessions, | Coming up at https://molecularwebxr.org/<br>(most likely https://molecularwebxr.org/handmol) | (coming up) |

| | interactive discussions, and concurrent work. | | |
|---|---|---|---|

## 2. Getting virtual molecules into the real world: moleculARweb for education in chemistry and structural biology using commodity AR

Our initial foray into democratizing immersive molecular experiences focused on what we term "commodity Augmented Reality", that is AR applications that run on readily available, everyday consumer devices such as webcam-equipped computers, tablets, and smartphones, without requiring any specialized headsets. This endeavor, inspired by the work and demonstrations made a decade earlier by Arthur Olson,[5] culminated in the creation of moleculARweb (https://molecularweb.epfl.ch), a free, open-source website hosting a diverse suite of AR web applications. These applications are primarily designed for chemistry and structural biology education, aiming to make complex molecular concepts more tangible and engaging.[27,36,37]

The core principle behind moleculARweb is marker-based AR user interface. Users print simple 2D fiducial markers (e.g., the "Hiro" and "Kanji" patterns, or two foldable cube designs, Figure 1A) from provided PDFs (3D printing with black and white plastic ink is also possible); then, by presenting these markers to the device's webcam, virtual 3D molecular models appear superimposed onto them and adapted to their positions and orientations in real-time (Figure 1B,C,E,F). This results in the virtual objects being controllable naturally by moving the fiducial markers, allowing full view of the objects by rotating the markers in space and allowing to move molecules in 3D, useful to for example explore how they interact (Figure 1B,C,E). This marker-based approach offers several key advantages for educational deployment. It works "out-of-the-box" in standard web browsers across various operating systems, requiring no software installation beyond what's typically available on modern devices. The primary hardware requirement is a device with a webcam and an internet connection, making it highly affordable and available. The workflow is straightforward: users navigate to a URL, present the markers to the camera in order to move the virtual objects around, using regular controls with a mouse to or touch buttons to execute actions. Manipulating the virtual objects with the physical marker provides a direct and tangible way to control the orientation and position of the virtual molecule, bridging the gap between physical models and digital representations. The fiducial markers can also be incorporated into texts such that students can visualize the molecules by scanning the markers with their phones, in which case they can then rotate the objects with touch gestures (Figure 1F).

Complementing the marker-based AR experiences, some of moleculARweb's activities possess a markerless mode that allows users to view the molecules "in their space" via look-through AR in smartphones and tablets. This mode leverages the spatial tracking capabilities of smartphones and tablets, fed into the app by the WebXR API, to display the virtual objects anchored to a plane detected by the system, allowing then users to freely move around the object to inspect it (Figure 1D and video linked to it, explanations in Figure 1H).

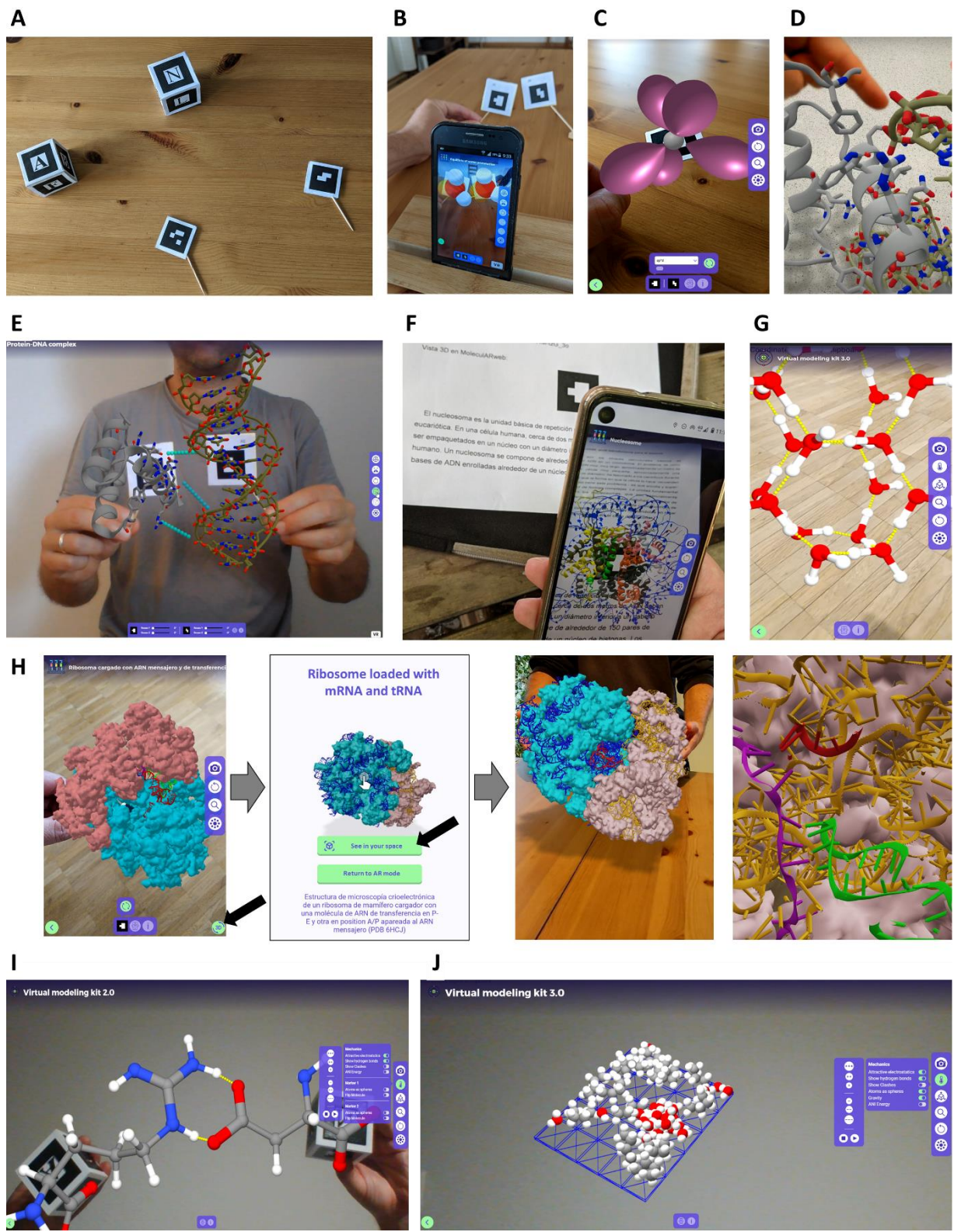

**Figure 1. Representative examples of moleculARweb's activities and playgrounds. (A)** The 4 main kinds of fiducial markers used in moleculARweb, created with a regular printer and folded into 3D cubes (markers for VMK 2) or glued to sticks for convenient handling (flat markers for all other activities). **(B)** The flat markers working in pass-through AR on a phone, here to test how two water molecules interact by hydrogen bonding and exchange protons. **(C)** An sp$^3$d hybrid orbital on a flat marker, from the activity on molecular shapes. **(D)** AR "in the user's space" without any markers, and the user pointing at a phosphate group in the DNA backbone of a protein-DNA complex (more in panel H). **(E)** Docking a protein molecule into a cognate DNA, guided by visual cues. **(F)** One of the regular flat fiducial markers, embedded in a text to show the 3D molecule on the phone (where it can be rotated by touch gestures). **(G)** A piece of ice loaded into the VMK 3.0 (more in panel J). **(H)** A 3D model of a ribosome displayed on a flat marker, and how to go from it into markerless AR: first tap on "3D" at the bottom right, then on "See in your space"; then scan a flat surface onto which the molecule will get anchored. (I) The VMK 2.0 being used in mirror AR mode to explore interactions between a positively charged arginine sidechain on the left and a negatively charged aspartate sidechain on the right; the app displays hydrogen bonds in real time (yellow dotted lines) because this is selected on the molecular mechanics pad. (J) The VMK 3.0 being used to show how polar and non-polar molecules phase-separate if polar forces are on, as controlled in the molecular mechanics pad.

*MoleculARweb's modules, activities and playgrounds*

In terms of content, moleculARweb is structured into thematic modules, each containing several activities. A module on *Atomic and Molecular Orbitals and Molecular Shapes* allows users to visualize *s*, *p*, *d*, *f*, and various hybrid atomic orbitals; it also includes activities for exploring idealized VSEPR (Valence Shell Electron Pair Repulsion) geometries and the molecular orbitals of simple molecules (e.g., $H_2O$, $CO_2$, $NH_3$) (exemplified in Figure 1C). Next, a module on *Hydrogen Bonding, Acids and Bases* allows students to interactively explore hydrogen bond formation, proton transfer equilibria (such as water autoionization or the reaction of an acid/base with water), and the dynamic nature of these interactions (Figure 1B).

Two modules on structural biology follow. *Atomic Structure of Biological Macromolecules* features visualizations of protein secondary structures (α-helices, β-sheets), protein-DNA complexes (including an activity where a protein and its cognate DNA are fitted by manual operations, Figure 1E), and related content. Then, with activities in the module on *Large Biomolecular Assemblies* users can explore complex structures like viruses, ribosomes, and nucleosomes, many of these activities supporting the markerless AR mode on compatible smartphones (example in Figure 1H).

At the top of the moleculARweb website, a special module of *Playgrounds* presents AR tools that behave as whole small programs rather than specific activities. The Virtual Modeling Kits (VMKs[36]) allows users to load any molecule via PDB coordinates that can be pasted into a box, fetched from databases, or drawn using an embedded JSME[41] applet based on hack-a-mol[42]. In the VMK 2.0 two molecules can be loaded, each displayed on a manipulable cube marker; each cube can then me moved independently to drive the molecules around as required to compare them, make them interact, etc. In the VMK 3.0 any number of molecules can be loaded to populate a box where they are contained. Both VMKs feature rudimentary molecular mechanics powered by the Cannon.js physics engine hacked to simulate thermal fluctuations and to emulate electrostatic forces that can be switched on and off at will. Additional controls allow users to switch between ball-and-stick and volume representations of the molecules, toggle display of hydrogen bonds, swap molecule chirality, emulate gravity (in the VMK 3.0, example in Figure 1G and 1J), and control the view (switching camera, cutting the background, etc.). While our dedicated papers explore several applications of both VMKs, we outline in the next two paragraphs some example use cases.

With the VMK 2.0, having the molecules attached to the markers allows exploration of their internal degrees of freedom and dynamics as the rotational and translational diffusion are removed (the molecule is re-centered at the cube in each frame), and driving interactions between pairs of molecules, where the app can show hydrogen bonds and atom-atom clashes. With these capabilities, teachers can show (and students can explore by themselves) how atoms rotate around single bonds, how non-planar rings exchange conformations, how aromatic and conjugated moieties remain planar, how molecules interact, etc. (example of the latter in Figure 1I). Another application of driving molecules with two cubes is in comparing stereoisomers, for example by loading the same molecule on two cube markers but inverting only one.

In turn, the VMK 3.0 behaves as a simulation box that allows teachers to show or students to explore how multiple water molecules stick together through polar forces to form a drop; one can even start with an example 3D model of ice to understand its peculiar water arrangement (Figure 1G) and how this structure is lost upon melting, in turn increasing the density. The melted piece of ice then behaves as a drop where the water molecules remain cohesive, yet some water molecules escape from time to time, illustrating vapor pressure that increases with temperature until eventually the system boils. Another interesting application uses the feature to toggle polar forces at will, to illustrate for example how miscibility and immiscibility work (Figure 1J).

The *Playground* module features also a marker-less commodity AR version of all the static content available in the website.[37] Built by using the Handfree library, in this web app users can set 3D virtual molecules, orbitals, etc and place them on the webcam's feed, to then move them with their hands.

Finally, the *Playground* module at https://molecularweb.org includes a web app called PDB2AR useful to create custom 3D material that can be visualized with the cube marker, with AR in the user's space, or in AR/VR headsets in stand-alone fashion or inside our website for multiuser molecular graphics MolecularWebXR. PDB2AR is treated in the section 3 below, and MolecularWebXR subsequently in section 4.

*User reception, usage statistics, and feedback*

MoleculARweb relies solely on standard client-side web technologies: HTML5 for structure, CSS for styling, and JavaScript for interactivity. 3D graphics rendering is handled by Three.js,

while marker tracking leverages AR.js, which itself is a wrapper for ARToolKit. This all makes the website entirely functional in regular laptops and computers equipped with a webcam, as well as in smartphones and tablets. The website is thus extremely easy to utilize, as attested by the high success documented by continuous access from all around the world (Figure 2A).

A very interesting application of moleculARweb reported from the University of San Luis in Argentina[43] shows how it can be integrated with other tools for AR visualization, in this case closing a cycle that connects chemistry with physics and biology by starting with two small molecules, then zooming out to the organismal system level and then zooming back in into molecules and orbitals (Figure 2B). As reported, the novelty brought by AR produces strong student engagement.

Accordingly, already with our first tests on high-school students and from initial online feedback, all right in the middle of the Covid-19 pandemic, extensive user feedback and web analytics indicated widespread global adoption and some positive pedagogical benefits that required deeper study yet clearly demonstrated a huge positive effect on student engagement.[25,27,44] In particular, the accumulated evidence from over four years of online teacher surveys confirms the capacity of moleculARweb to make difficult chemistry and structural biology topics more approachable and engaging. Educators repeatedly highlight how the platform helps simplify concepts like molecular bonding, geometry, and macromolecular assembly, turning abstract topics into interactive, manipulable experiences. Teachers appreciate not only the pedagogical value of the activities but also the platform's flexibility, being able to integrate custom content tailored to their curriculum, pace, and students' needs.[45,46] Such adaptability allows the same activity to be used differently in a high school chemistry lesson, an undergraduate biochemistry class, or informal science outreach. In addition, teacher testimonials consistently point to increased student enthusiasm and participation when activities include hands-on AR interactions—students who might be passive during traditional lessons become actively involved in exploring and discussing 3D structures. However, feedback has also illuminated areas for improvement, that we tried to attend as it becomes possible—with many of the requested features actually available already, but documentation gaps mean some users remain unaware of them.

From a broader perspective, moleculARweb exemplifies both the promise and the ongoing challenges of integrating AR into STEM education. Immersive tools reliably capture attention and spark curiosity, yet engagement alone does not guarantee deeper or longer-lasting learning. The novelty of AR may be a key driver of student interest today, but sustained educational value depends on integrating these tools into pedagogically sound activities that actively support comprehension, retention, and skill-building. Furthermore, although anecdotal reports suggest AR can accelerate spatial understanding of molecular structures—freeing instructional time for higher-level discussion—systematic, controlled studies remain scarce. These considerations point to an important next step: pairing widespread deployment of platforms like moleculARweb with rigorous, longitudinal research that evaluates learning gains, not just engagement. This would involve interdisciplinary collaboration between developers, educators, and education researchers to design studies that measure outcomes such as conceptual understanding, retention over time, and transfer of spatial skills. Ideally, such research should compare AR-based instruction to both traditional 2D approaches and physical model use, and should investigate how factors like prior knowledge, spatial ability, and content complexity influence effectiveness. In the meantime, the teacher and student feedback collected already offers us a roadmap for iterative improvement: enhance discoverability of existing features, expand the content library, refine interaction methods for mobile users, and explore integration with complementary tools (e.g., simulation engines, AI-based assistants). Addressing these points, while preserving moleculARweb's core strengths of accessibility and ease of use, could further solidify its role as a bridge between fascination and meaningful learning in molecular sciences education.

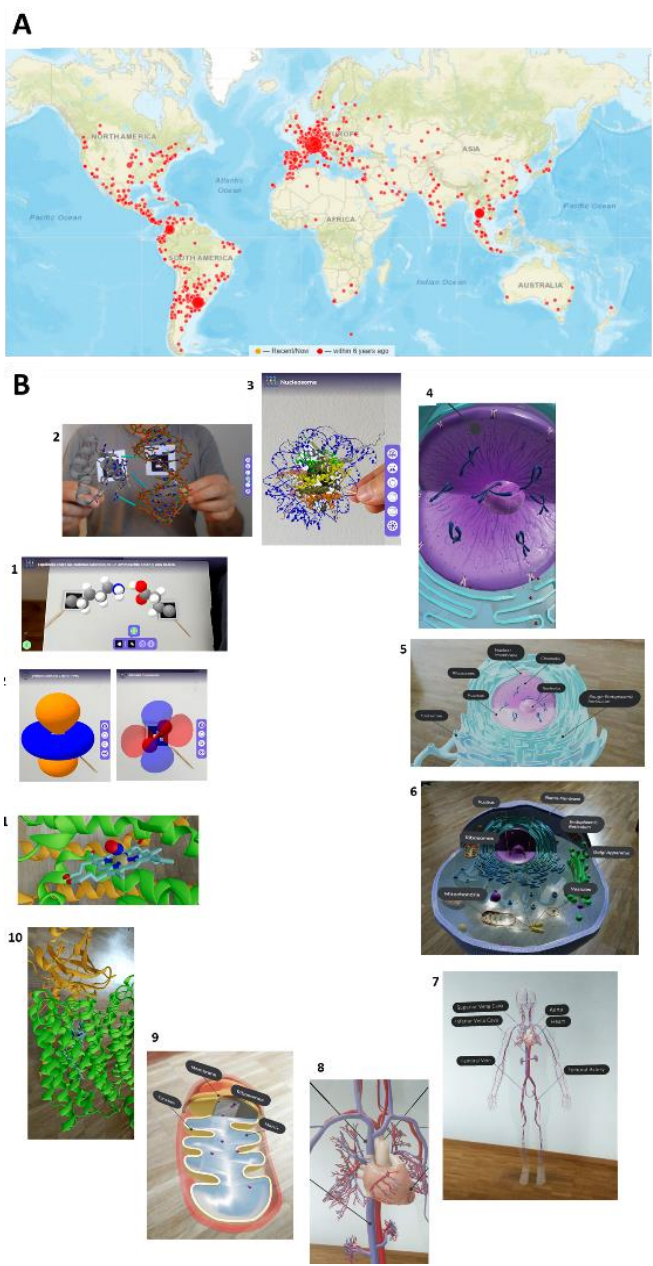

**Figure 2. moleculARweb usage statistics and example integration with other AR tools.** (A) Accesses since launch, as tracked at https://clustrmaps.com/site/1ar2f. (B) Integration of moleculARweb bringing AR views of chemistry and structural biology with Google-based AR views of systems, organs, tissues, cells and organelles. This is part of an activity connecting physics, chemistry and biology, all in AR, starting from electrostatic interactions (1) relevant in biology for example for protein-DNA complexation (2) and compaction into nucleosomes (3) and chromosomes (4) inside the cell's nucleus (5). A full view of the cell (6) then leads to the systems level (7), and then the activity zooms back in going through the heart as an organ (8), a mitochondrion (9) where cytochrome c oxidase (10) carries out that last step of respiration at its heme group (11) where oxygen interacts with a specific orbital with a specific orbital at the heme's iron site (12).

## 3. The PDB2AR tool for custom content creation for all devices

An important aspect of democratizing these immersive technologies is to empower educators, researchers, and science communicators to create their own customized content without requiring extensive programming expertise or specialized software development skills. We experienced this urge first-hand when a few months after moleculARweb's release several teachers had asked about the possibility of adding specific content. To address this need, we developed an online utility integrated within moleculARweb's *Playgrounds* module, named PDB2AR (or "Build your own webAR/VR views", Figure 3).[38] This tool, accessible directly at https://molecularweb.epfl.ch/pages/pdb2ar.html, provides a streamlined workflow for generating shareable AR and VR experiences that work on the cube marker (in all devices), in the user's space via WebXR if running in a smartphone, or in full VR if running on the web browser of a AR/VR headset. Besides, the virtual objects created with this tool can be placed in sessions of our MolecularWebXR platform, described in section 4.

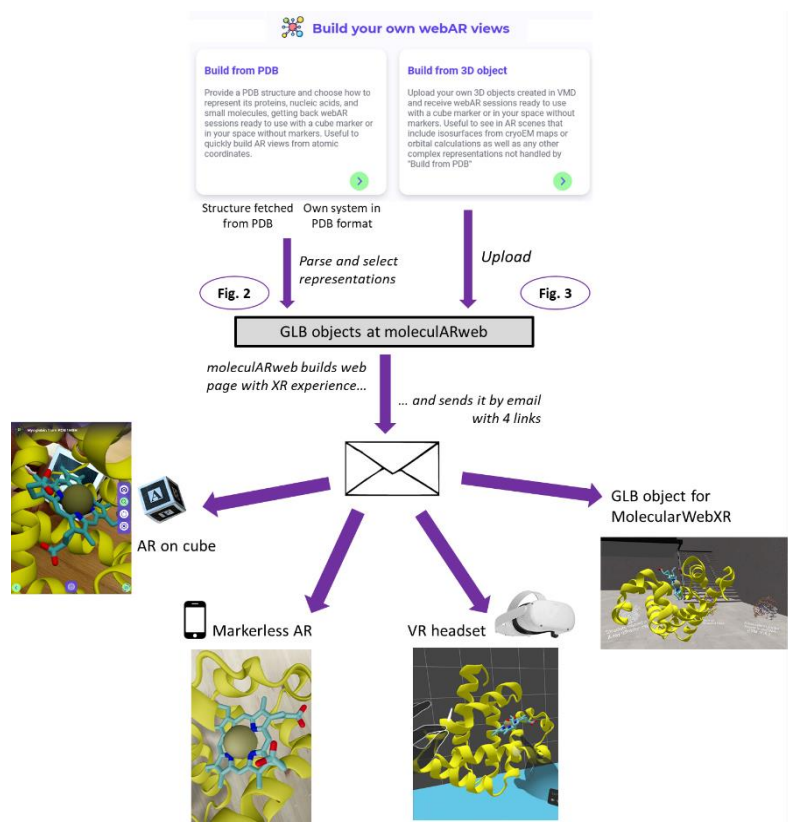

**Figure 3. Custom content generation within the moleculARweb universe with PDB2AR.** Starting from a PDB file that will be parsed by the app and converted into the user-selected representations, or from wavefront objects exported from VMD, PDB2AR creates AR/VR experiences that run using the cube fiducial marker, markerless AR "in the user's space", or in headsets, also providing validated links to add the elements into MolecularWebXR scenes.

*Using PDB2AR*

As input to PDB2AR, users can upload a PDB file containing atomic coordinates from their local system, fetch a structure directly from the Protein Data Bank (PDB) using its unique PDB ID, or retrieve a model from the EBI's AlphaFold Protein Structure Database using its UniProt entry. Alternatively, users can load virtual objects generated with VMD (exported as WaveFront OBJ+MTL files); this allows them to visualize not just molecules but also isosurfaces, grids, volumetric maps, etc., and thus essentially any other 3D element visible in their starting VMD scene (such as electron density maps, molecular orbitals, and custom geometric objects).

When a PDB file is provided as input, the web interface parses its molecules and then users can select how different molecular components (e.g., protein chains, nucleic acid strands, small molecules, ions, other molecules) should be displayed. Options include various standard representations like cartoons, sticks, spheres, surfaces, etc., along with a choice of color schemes (e.g., by chain, by element, by secondary structure); all following conventions of the VMD software.[47] In fact, PDB2AR will then generate a VMD script that will generate OBJ+MTL files, and the user can optionally edit this script in order to finely control the generated object.

Once the input and display options are set, the user provides a title for their scene and an email address. The PDB2AR tool then automatically processes the input on our server, generating the 3D virtual model in GLB format and with a special compression that makes the files light. The user then receives an email with various unique links to web pages that display the generated object in various forms, that is either anchored to the standard moleculARweb cube marker, or in the user's space in pure augmented reality, or as a virtual object; besides, a unique link to the GLB file allows seamless addition of the element into a MolecularWebXR scene

As we see with a number of submissions per month, the PDB2AR tool significantly lowers the barrier to entry for creating custom immersive molecular content. With it, educators can tailor visualizations precisely for their specific teaching objectives and researchers can quickly generate VR-ready models for collaborative discussions, presentations, or personal exploration of complex structural data. To us, PDB2AR has been instrumental in expanding the content available at moleculARweb and MolecularWebXR, as required for numerous educational and science communication activities.

**4. Deeper immersion without compromising reach, with WebXR: MolecularWebXR**
The maturation of the WebXR API presented a compelling opportunity to extend our web-based tools for molecular graphics into more deeply immersive AR and VR environments. The goal was to retain the core advantages of web delivery (no installation, broad accessibility) while offering a richer, more engaging and especially more immersive experience, that is also natively multiuser as all users in a session can see each other's avatars and talk naturally in the room as they manipulate virtual objects, possibly physically together or separated by thousands of kilometers. This is materialized in MolecularWebXR (https://molecularwebxr.org, Figure

4A), a platform designed for multiuser discussions, educational sessions, and collaborative exploration within virtual "rooms" or "sessions" where users connect together (Figure 4B,C) to concurrently view and manipulate models of relevance to chemistry, materials science and structural biology, that is not just molecular structures but also representations of molecular surfaces, orbitals, electron maps, etc. Each user can experience the session immersed in VR or AR with headsets (Figure 4D), in pass-through AR or cardboard goggle-based VR with smartphones (Figure 4E shows the latter) or in non-immersive way with every device.

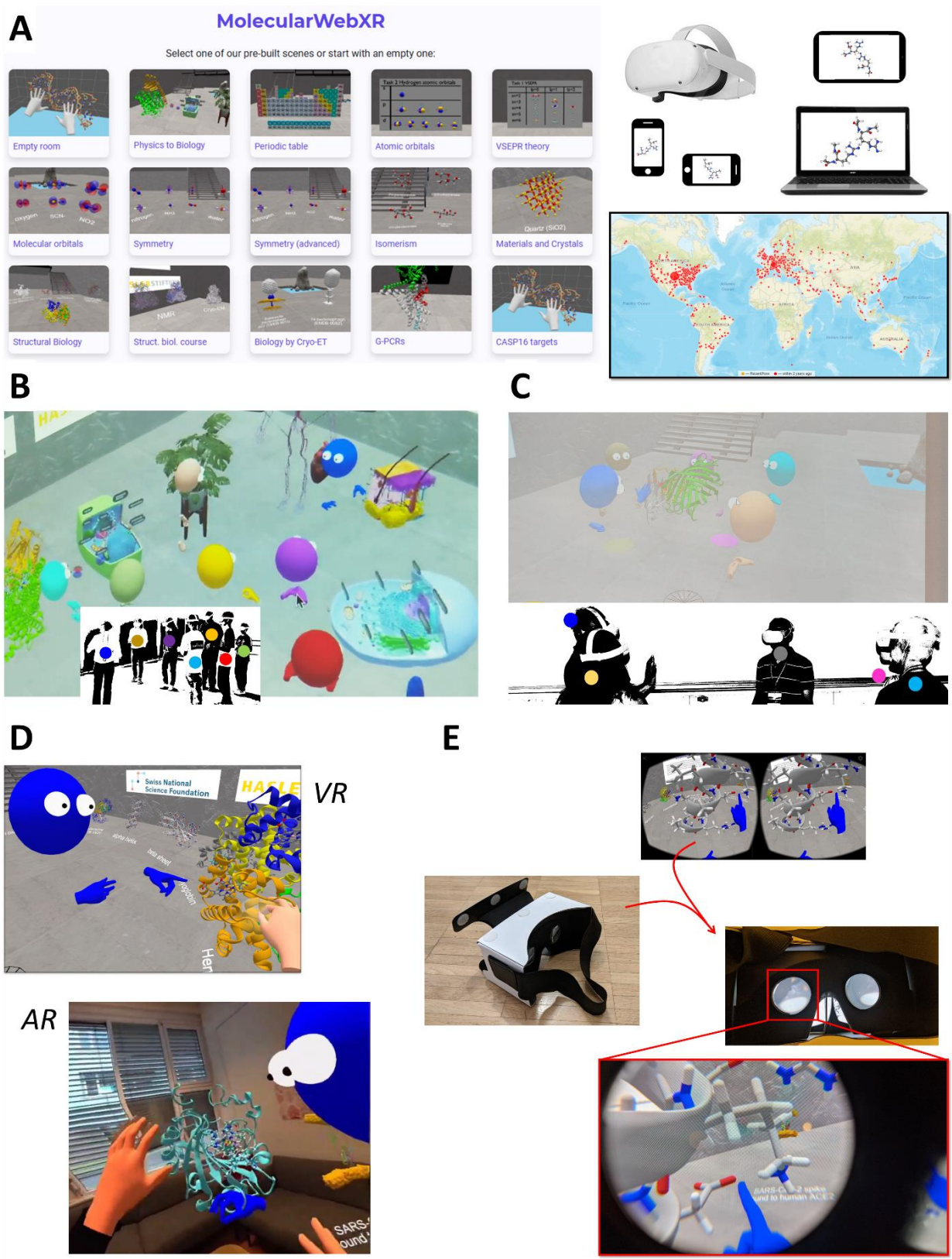

**Figure 4. A summary of MolecularWebXR.** (A) The rooms with predefined content available as of mid-2025, filled with objects created by us and collaborators using PDB2AR plus others obtained for free or purchased at Sketchfab.com. The map in the inset shows worldwide access since release. (B) A public communication session where a person (blue avatar) walks 6 guests around a virtual room connecting physics with chemistry and biology. Notice the presenter's hands pointing at sections of a model of a dermal tissue. Photograph taken at EPFL's Portes Ouvertes event in 2023, Lausanne. A short video demonstrating how this works is available at https://www.youtube.com/watch?v=NnKq4L3NVlY. (C) Five scientists discussing a protein structure; notice the presenter in blue pointing at the interface between two structures; photograph taken at the closing conference of the NCCR TransCure program in 2022, Bern. (D) Top: MolecularWebXR in VR mode, view of the user with the salmon avatar (notice the hand at the bottom right). Bottom: MolecularWebXR in AR mode, view of the user with the orange avatar. (E) MolecularWebXR experienced in immersive VR on phones, by tapping VR and plugging the device into carboard goggles equipped with lenses.

*Accessing and using MolecularWebXR*

To access the platform, free and without registration, users must direct the web browser of their device of choice (AR/VR headset, smartphone, tablet, computer) to https://molecularwebxr.org. Note however that the MolecularWebXR website available as of early 2025 will become the "XR Science" branch of the website starting late 2025 or early 2026.

To use MolecularWebXR, an admin host must first create a session starting from a room with preset content or with an empty room to be populated with objects built with PDB2AR. The created session will have unique identifiers (4-letter codes under "Invite people") that the admin must send to those that should connect to it via the "Join room" button at the landing page. The admin user has access to several features to control the session, such as removing or adding objects, changing their sizes, resetting them, controlling which users can grab objects and what objects can be grabbed, enabling or disabling audio, etc. For a more comfortable use of these features, it is best to create the session on a computer.

As users join the session, they can enter AR or VR or simply stay outside of XR, depending on the device's capabilities (examples of the different modes in Figure 4). Of course, the session is maximally experienced through AR/VR headsets, which offer the most immersive experience. Accessing with headsets, users are represented as avatars composed from their heads and hands or controllers, continuously updated as they walk and move their limbs around the virtual space. Users wearing headsets can interact with molecular models using built-in hand tracking or handheld controllers. By pinching with the index and thumb fingertips of one hand on a model (or pressing with the index finger on the controller), the user will grab and move the object. By pinching or pressing an object on two points, the user will change the object's size. Users can also talk with each other directly through the headset, a feature available in all devices that the host must disable if the users are also physically together.

Thanks to its web-based nature, MolecularWebXR also allows users to join sessions from smartphones and tablets, where they can navigate the 3D space using on-screen joysticks or touch gestures, and observe the avatars and actions of users who accessed with headsets. Moreover, they can also go into immersive VR by using cardboard goggles for stereoscopic VR view (Figure 4E) or in passthrough AR with 6 degrees of freedom. And of course the website also works in regular computers, where participants can use keyboard and mouse moves for navigation and viewing.

*Content available at MolecularWebXR*

MolecularWebXR includes several pre-defined virtual rooms containing curated content relevant to specific educational topics on chemistry, structural biology and materials science. Since this content evolved substantially since our paper presenting the website, we dedicate here quite some detail to all the content available as of mid-2025 (Figure 4A).

The first room, an "Empty Room", provides a virtual space without any objects, to be populated with material prepared with PDB2AR by clicking on the Elements menu and creating new ones by pasting the URLs to GLB objects received from PDB2AR by email. Next to this empty room, "Physics to Biology" is at the core of a science communication story connecting biology with chemistry and physics, described below in a specific section. Then, 8 rooms with material for introductory chemistry follow: a periodic table where the elements can be taken out one by one, useful for interactive quizzes and other activities (an XR version of the activities developed by Limpanuparb et al[48]); a room with the atomic orbitals and another with the HOMO and LUMO molecular orbitals of molecules selected for pedagogic interest; a room on VSEPR theory; two rooms on molecules symmetry, one passive and one where students must grab objects representing symmetry axes and planes and use them to identify the elements in various molecules; one room on isomerism; and one with example structures of materials and crystals. Finally, 5 rooms are dedicated to different aspects of structural biology: a room that introduces protein and DNA structures; one used at a structural biology course and covered in more detail below; one on cryo-electron tomography covering very modern and interesting 3D views of viruses, cell interiors and other structures; one on GPCRs focusing on its activation mechanics; and one with interesting targets from CASP16.

*MolecularWebXR case 1: Communicating science "from physics to biology"*

The room called "Physics to Biology" is at the core of a science communication story connecting biology with chemistry and physics that we carry out with a presenter and up to 4-6 visitors, all inside VR with the headsets set to match the virtual and real environments so that all the people inside the session can move around freely. Overall, the presentation covers the same content as that presented in Figure 2B but in a much more immersive and interactive fashion, as seen in the photograph in Figure 4B. The room contains a model of a human body's interior, where the story starts, followed by a model of a heart, a piece of dermal tissue, and an animal cell through which the story continues. Next to the human cell and almost attached to its membrane, there is a tiny model of a SARS-CoV-2 virus whose interior can be inspected upon zooming. Back to the cell and focusing on the nucleus and its chromosomes, the story continues down the line of a chromosome that unfolds slowly to reveal how DNA is compacted, including along the way a crystal structure of a histone nucleus with two turns of DNA wrapped around. Towards the most uncompacted region of the DNA, an X-ray structure of a transcription factor bound to DNA allows the story to present the shape of the DNA molecule with its minor groove and major groove, the bases stack and interacting with each other via hydrogen bonds, the transcription factor's alpha helix docked into the DNA's major groove with some unspecific salt bridges to the backbone phosphates and some specific interactions to the bases.

After talking in such detail about the X-ray structure of the transcription factor bound to the DNA molecule, the story goes on with a piece of an X-ray structure where the electron density map is overlaid, serving to talk briefly about how scientists get to know the structures of molecules. The story then briefly refers to the structure of human carbonic anhydrase bound to a sulfonamide-based inhibitor, as an example of how small molecule therapeutics work.

Then, back to the start, a model of a plant next to that of the human body sets the stage to talk about plant cells and their differences with animal cells. When talking about the thylakoid the story can move on to the 3 oxygen molecules leaving the plant cell, each with a different molecular orbital being shown: sigma and pi bonding orbitals, and the anti-bonding orbital with which oxygen binds the orbital protruding out of the heme group in cytochrome c oxidase. The structure of the oxidase is shown, and the storyteller can grab the oxygen molecule and show how it "diffuses" into the oxidase and towards the heme group, where it will finally accept electrons at the end of the respiratory chain to become water.

*MolecularWebXR case 2: Illustrating a course on structural biology*

The room "Structural Biology Course" was designed to support a course focused on protein structure, protein dynamics, and the various experimental techniques used to explore them. Like in the previous case, in this room a presenter talks with up to 4-6 visitors, all inside VR with the headsets set to match the virtual and real environments, a story devised to match the content.

In this case the story opens with the same animal cell model as the above-described room, and it follows with the same models of the unfolding chromosome together with the nucleosome and protein-DNA complex (Figure 5A). But then once the structure of the protein-DNA complex is explained, the story moves on to the detailed structural and dynamical features of proteins, in the framework of the techniques utilized to determine them.

First of all, the interactions that held together alpha-helices and beta-strands are covered in detail with models that include the hydrogen atoms and thus allow to very explicitly show the hydrogen bonds at their cores (Fig. 5B). This can be accompanied by inspection of the other models, to go to quaternary structure; more content about secondary, tertiary and quaternary structure is available in the room titled "Structural Biology".

Next in the story, 4 models of malate synthase G cover the 3 main techniques for structure determination at atomic resolution, that is NMR, X-ray diffraction and Cryo-EM. Malate synthase G is the only protein whose structure was solved with all three techniques, representing the high-size limit for NMR spectroscopy and the small-size limit for Cryo-EM. Two representations are shown for the NMR structure; one displays in green a few examples of the long-range NOEs used to determine the structure, and the other model shows members of the ensemble in cartoon representations colored by the RMSD to the mean structure. In the ensemble, a highly flexible loop is shown that provides a chance to explain how this might be due to real dynamics and/or to a lack of assignments and restraints to define this region's details upon structure calculation (Figure 5C). Moving on to the Cryo-EM and X-ray structures, the story explains how that same region is not modelled at all in the X-ray diffraction structure, and it is modeled despite missing density in the Cryo-EM structure. The X-ray structure is colored by C$\alpha$ B-factors, which provides a chance to identify other somewhat flexible regions and compare them with the NMR ensemble.

To conclude, the session also covers the same sulfonamide-bound human carbonic anhydrase covered by the room connecting physics to biology, and also the structure of the ABL kinase with imatinib bound to it. At the level of the structural biology course, the story can go deeper to show for example how deep imatinib goes into the kinase, pointing at some key contacts, and detailing the structure around human carbonic anhydrase's zinc ion.

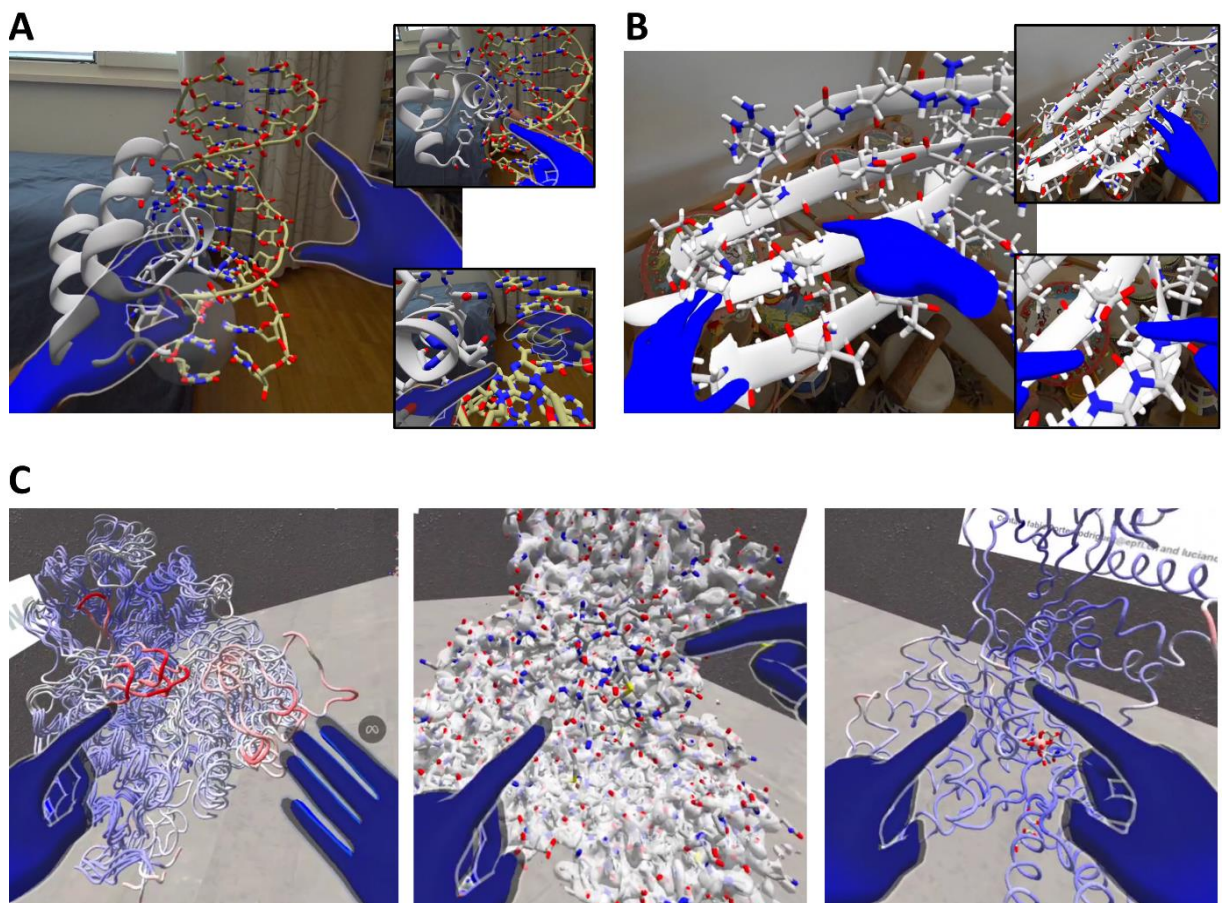

**Figure 5. MolecularWebXR's room presenting a course on structural biology.** (A) A user showing with his hand the major groove of a dsDNA molecule, where a helix of the transcription factor docks. The minor groove is also easily spotted in this 3D model, as well as the non-specific salt bridges between DNA's backbone phosphate groups and positively charged sidechains of protein residues (top inset), and the hydrogen bonds between protein sidechains and DNA bases (bottom inset). (B) A user pointing at one of the hydrogen bonds that stabilizes pairing of beta strands in a beta sheet. In this model it is also easy to spot charge-charge interactions across strands (bottom inset), and how one side of the sheet facing the interior of the full protein is hydrophobic (top inset) while the solvent-facing side is very polar (main panel). (C) Structure of Malate Synthase G as determined by (from left to right) NMR, Cryo-EM and X-ray diffraction; in the three cases pointing at highly flexible loops that serve to explain students how each technique works, what they can reliably solve, and more. In particular, the loop on the left lacks density in X-ray diffraction maps hence it is not modeled, lacks restraints in NMR hence appears very flexible in the ensemble, and has very low density in the Cryo-EM map.

*User reception, usage statistics, and feedback*

While the molecular models within these rooms are static in terms of their internal structural dynamics (unlike the dynamic simulations in the VMK presented in Section 2 or HandMol presented in the next section), the platform excels at providing shared, immersive viewing and discussion spaces. We have applied this tool at science communication activities as shown with use case 1 in Figure 1B, at bachelor and doctoral level courses, and at scientific conferences as in Figure 1C. MolecularWebXR is featured by Meta's web browser, was nominated to a PolyXR award in 2024, and is today widely access as shown by statistics (inset in Figure 4A).

## 5. Towards interactive, immersive, multiuser molecular graphics and modeling: HandMol

With large numbers of users around the world, moleculARweb and MolecularWebXR provide accessible web-based tools for immersive molecular visualization and shared immersive discussions. However, these tools primarily handle molecules as static 3D objects, or in the best case such as in moleculARweb's VMKs with some flexibility but limited capabilities for immersive graphics and fine 3D manipulation. The ambitious next step was to move beyond towards truly intuitive, dynamic, physics-informed and multiuser manipulation. We wanted to empower users to not just see a molecule in immersive multiuser 3D as possible with MolecularWebXR, but also to grab it, bend it, explore its energy landscape, and even simulate simple reactions with intuitive, bare-handed interaction. This vision required a fundamental architectural shift: instead of loading pre-rendered 3D models (GLB files), our application needed to treat molecules as collections of individual atoms and bonds, each capable of being rendered, processed and displaced separately, as in moleculARweb's VMKs but with MolecularWebXR's XR capabilities. This goal led us to web apps for genuine single- or multi-user molecular graphics and modeling with option for immersive graphics, giving rise to the HandMol series of apps that consist of a client-only prototype[40] and a MolecularWebXR-based HandMol app, the former already available as a stand-alone web page without support and the latter coming up in late 2025/early 2026 in a stable release merged with the whole MolecularWebXR platform.

### *The client-only HandMol prototype as a proof of concept*

Our first implementation of the overarching idea was the HandMol client-only prototype, a system designed to validate several components required to couple WebXR with AI and molecular simulation engines. This prototype, available at https://go.epfl.ch/handmol, employs a two-part architecture where two web applications work in concert, connected wirelessly via the WebRTC protocol without any intermediate backend servers (i.e. being fully client-side except for the initial WebRTC connection).

The two web apps that make up the client-only HandMol open at different links within the main page: HandMol-VR and HandMol-Computer. HandMol-VR is meant to run in the web browser of a headset to provide immersive AR or VR experiences where users can grab and displace atoms at will, as well as get information about atom names, distances, etc. Like in MolecularWebXR, AR and VR can also be experienced, of course without hand tracking capabilities. In turn, HandMol-Computer is meant to run on the web browser of a computer that will be used to move files easily into and out of the session, but it can also run in a second headset to support two-user immersive sessions.

The client-only HandMol prototype successfully integrated a suite of modern technologies to create a uniquely interactive experience. Three.js and the WebXR API provide the immersive graphics and natural interaction capabilities. Meanwhile, the Cannon.js physics engine running inside the web browser provides real-time feedback on molecular flexibility and atomic clashes similar to those in moleculARweb's VMKs, giving a tangible feel to manipulations. For more accurate simulations, the system can offload a given set of atomic coordinates to external APIs that perform energy minimizations, after which Cannon takes control back. Two such APIs are available, one running the ANI-2x neural network potential for fast, high-quality calculations on organic molecules,[49–51] and another running the AMBER14 force field (via an OpenMM backend)[52] that works on standard biomolecules. Energy minimizations can be triggered at will or automatically upon releasing an atom after moving it, providing immediate, physically realistic feedback.

The prototype also implements a rudimentary way to handle human-computer interaction through an LLM, in this case GPT-3.5-Turbo for which the user must input an API key. Running in a Chrome browser of a computer, the HandMol-Computer app can use this browser's Speech Recognition API to capture voice commands in natural language (e.g. such as "hide the hydrogen atoms" or "show carbon atoms in magenta"). The LLM parses this command and translates it into a function call within the application, freeing the user from learning any menus or scripting languages. While the scope of this LLM test is narrow, it served as the basis for more comprehensive natural control of the MolecularWebXR-integrated HandMol program.

This prototype robustly demonstrated that highly interactive, AI-assisted, and computationally-grounded molecular modeling was feasible entirely within a web-based framework, with user tests (detailed in) showing significant gains in intuitiveness and task completion speed for complex 3D manipulations. More details available in a dedicated preprint[40] and in a blog post that discusses implementation of the key technologies including pieces of explained source code.[53]

***The MolecularWebXR-integrated HandMol: from regular but easier to use molecular graphics to immersive interactive simulations and multiuser discussions***

Building on the success of the HandMol prototype, the next evolution is the MolecularWebXR-based HandMol. It is intended as a regular molecular graphics program, i.e. where a molecule is seen against a black background and controlled with mouse gestures, but with the capability to spin up all those technologies tested in the HandMol prototype, and more. First of all, this (upcoming) web app can go immersive in AR or VR, again with head-mounted displays or on phones or tablets or as a first-person role game in computers (Figure 6A). Second, multiple users can connect to see and act on the same molecular system, each from its own viewpoint regardless of what device they use to access (Figure 6A). Third, users can interact with atoms or beads of simulation by pinching atoms, while they can see atom names and atom-atom distances by pointing at them with index fingers (Figure 6C). Fourth, actions and views can be controlled through a menu displayed inside AR/VR from the user's left wrist (Figure 6B), with regular menus as in traditional molecular graphics software (available to the Host only, as in Figures 6D and 6E), and by typing or speaking up requests in natural language (Figure 6D; note that the speech recognition feature is only available for the Host and when using Chrome web browsers). Figure 6E covers the controls and options available in the main box menus.

Currently at an advanced stage of development, HandMol is set to be integrated into the main MolecularWebXR platform as a production-ready public service in late 2025/early 2026. This new HandMol is not a two-part application but rather a single-web page application that works like MolecularWebXR, that is where a Host creates a session—in this case starting from atomic coordinates in PDB format—and shares the session's ID for its own or other users' access through immersive (or any regular) devices. This integration represents the culmination of our efforts, bringing dynamic simulation into the same space as our multi-user presentation and education rooms. Like in the previous apps, this again supports AR or VR, in headsets or regular devices, and for single or multiple users.

The architecture of the new HandMol is more robust and scalable than that of the HandMol prototype, using web sockets to centralize virtual sessions and synchronize them with molecular simulations running in a server. When a Host user creates a session in the new HandMol, the server spins up a dedicated container for that specific session, which runs the chosen simulation engine to support interactive MD. Available MD engines as of the date of release will be the fast rigid body-based system based on the Cannon.js library (derived from moleculARweb's VMKs[27,36] and exemplified in Figure 6F), the ANI-2x AI system[49] or OpenMM-based AMBER14 (with automatically generated parameters or with user-provided parameters) for atomistic simulations with higher physical accuracy, and Calvados[54] for coarse-grained simulations of proteins (inserted in membranes if chosen, example in Figure 6G). One special container, called simply HandMol, provides visualization-only capabilities (with all other features enabled).

In the new HandMol, the XR sessions synchronize all users with each other and with the simulation engines via a persistent socket connection that requires high bandwidth and computing power. Therefore, the backend manages resource allocation preventing overload by tracking server load and limiting the number of concurrent computer-intensive simulations, and we have opted to limit each session to 15 minutes—those with special needs for longer sessions can contact us directly to discuss options.

On the client side, the application retains and refines all the core functionalities of the prototype: intuitive bare-hand manipulation, on-the-fly molecular mechanics feedback (where energy minimization was replaced by a continuously running simulation that can work as a continuous minimizer by setting its temperature to zero), and a LLM-powered voice/type assistant much improved from the original prototype, in this case powered by OpenAI's GPT-4o-mini. Furthermore, users can use in-VR menus to dynamically adjust parameters like simulation temperature or zoom level. The system also visualizes forces and interactions through "rubber bands" that appear when a user pulls on an atom. Additionally, pointing at an atom with the index finger tip displays the atom's name, and when two atoms are pointed the distance between them is displayed as well (Figure 6C).

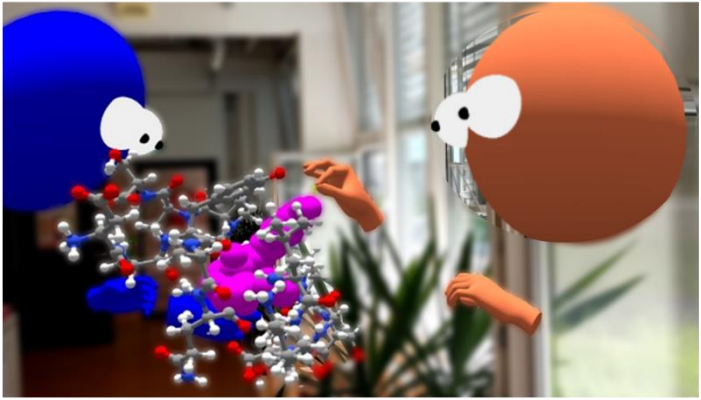
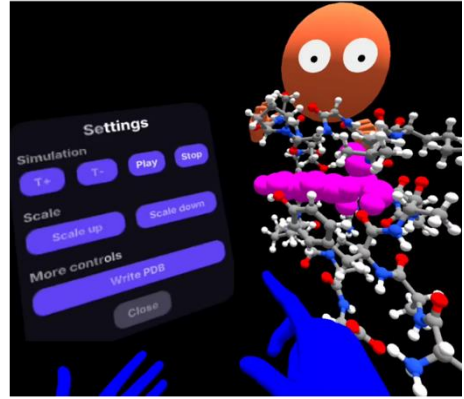
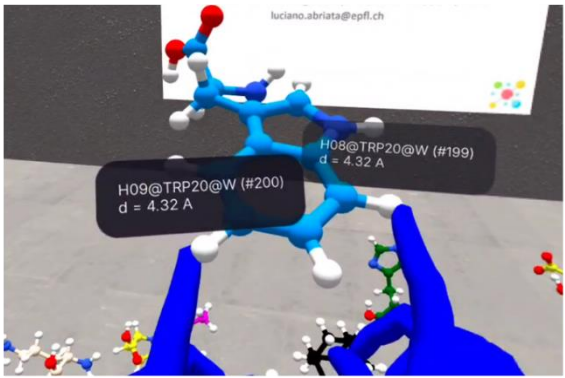
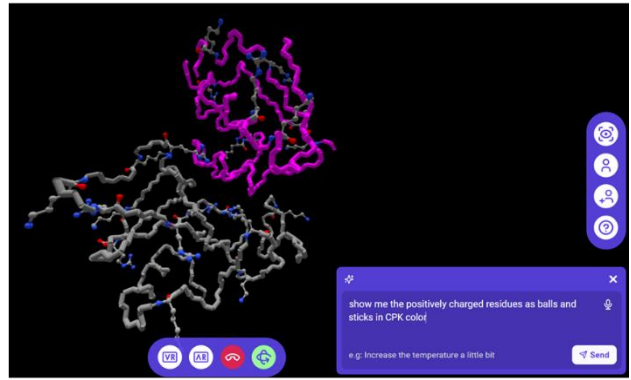
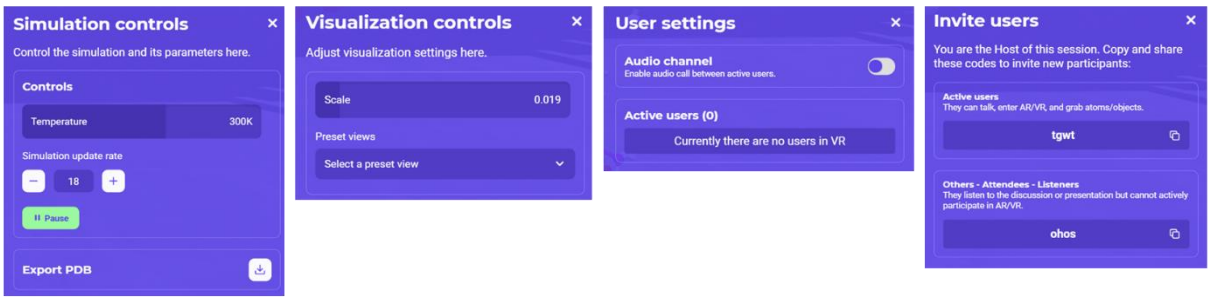
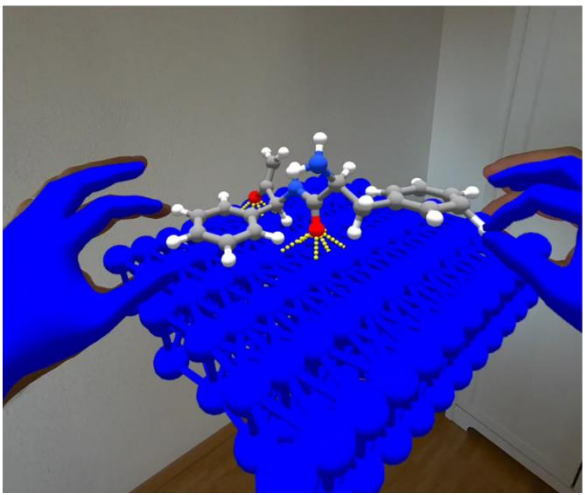
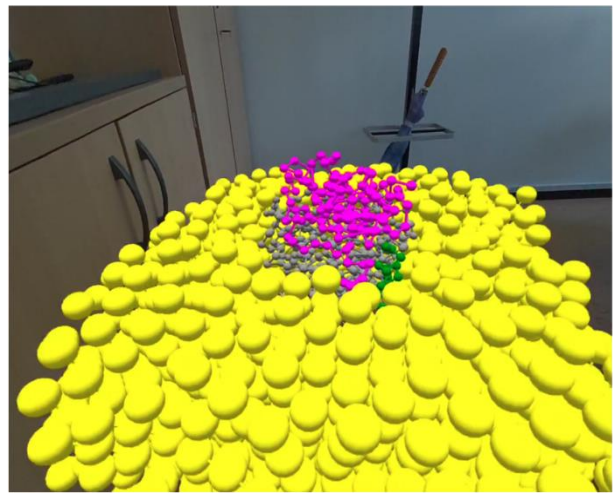

**Figure 6. Some features of the upcoming MolecularWebXR-integrated HandMol web app.** (A) Two users acting on a molecular system in XR, as seen by a third one who accessed through AR on a smartphone (the background was blurred upon figure preparation, for clarity). (B) Same session as in A but seen from the viewpoint of the user in blue avatar. This user has unfolded a menu accessible on the left wrist, with quick commands that can be executed right from within the immersive experience. (C) On pointing at an atom, its name is displayed with format Atom@Residue@Chain (atom number). When two atoms are pointed at, the distance between them is shown under their names. (D) The web app being used as a regular molecular graphics program by a single user on a laptop, controlling the molecule's orientation with the mouse ("orbit controls" on). This panel also exemplifies spoken commands being recognized (AI assistant on the bottom right). (E) Overview of the box menus available to the Host: *Simulation* controls allow to start and stop simulations and control the temperature, as well as downloading in PDB format the coordinates as currently laid out. *Visualization* controls allow quick scaling of the molecular system and provides access to quick preset views. Under *User settings* the Host can see who is connected to the session after they have been invited with the codes displayed in *Invite users*. The AI module works separately and is shown in action in panel D. (F) The rigid body-based MD engine can handle all elements from the periodic table plus some special-purpose atoms (HX, CX, NX, OX as color alternatives to the regular H, C, N, O atoms) and coarse-grained beads (under development); besides, specific interactions can be hard-coded into the PDB. In this example, a bilayer of Cu atoms is set to interact with O atoms, which helps the user to dock a molecule of biphenylalanine in preparation for subsequent studies via for example DFT calculations as in.[55] (G) Display of systems simulated with the CALVADOS engine utilizes specialized beads for the 20 amino acids plus two types of beads for the membranes. Shown in this panel is the x-ray structure of *T. thermophilus* cytochrome *c* oxidase (3 polymer chains) inserted in a membrane mimic. All the features shown in this figure are at the moment evolving and might look different when the actual first version of the platform is out. For more examples, including some of HandMol running in visualization-only mode or interactive MD with Amber and ANI engines, see Figure 7.

## *Example use cases of the HandMol web apps*

By integrating dynamic, physics-based modeling directly into the multi-user MolecularWebXR platform, HandMol achieves our goal of creating a single, web-based ecosystem that spans the entire spectrum from simple AR visualization to shared, immersive discussions, and finally to collaborative, real-time, AI-assisted interactive molecular simulation—all in an accessible format that is just an URL away.

We have so far used our programs to get a "feel" on the strength of certain interactions, to manually dock small molecules into pockets, to explore possible conformations and effects of chemical modifications, to setup systems for more thorough downstream analyses using regular molecular simulations, to teach about molecular structure, and to discuss molecules. In this section I present specific use cases, some illustrated in Figure 7, moving beyond applications to education (showing just two examples: Figure 7A and 7B) by exploiting actual applications to our research work (Figure 7C-F plus utilization inside the cited works).

Through mechanics assisted by the rigid-body dynamics engine, we have used the client-side HandMol prototype and the upcoming MolecularWebXR-integrated HandMol to dock small molecules into proteins (example in Figure 7C). We found this of utmost utility when clear sources of information and physicochemical patterns define pretty obvious poses but regular docking programs do not produce compatible complexes. For example, we used the rigid-body engine to manually dock substrates into the two active sites of a newly discovered enzyme based on templates of similar small molecules bound to similar domains, to then relax the complexes by atomistic MD and derive residues studied by mutagenesis.[56] Similarly, we used interactive simulations to thread single-strand DNA molecules into a protein nanopore (Figure 7D) and to place small molecules inside protein nanopores,[57] in both cases for downstream studies through extensive, regular MD simulations. Notably, both tasks are very difficult to achieve with regular 2D software and without the assistance of a forcefield, especially when threading the bulky DNA strand into the nanopore; here is when interactive MD with the modern HandMol displays its main power.

With interactive simulations driven by an Amber forcefield, we found the upcoming HandMol useful to explore protein mechanics hands-on and very intuitively. For example, we tested how a protein nanopore might collapse under strong electric fields and other sources of destabilization,[58] we probed how a transcription factor interacts with DNA (Figure 7E), and we modeled how a disordered peptide might close up on itself upon phosphorylation (Figure 7F). During such explorations, it appears in our hands that seeing the yellow strings stretch and compress like rubber bands convey a visual feel of the forces applied, consistently with two studies showing that interactive MD in VR allows human subjects to sense molecular properties purely through visual cues as they interact with them in VR[59,60]—making haptic feedback, which is currently very expensive and hard to integrate into devices of mass use, less needed for a full immersive experience.

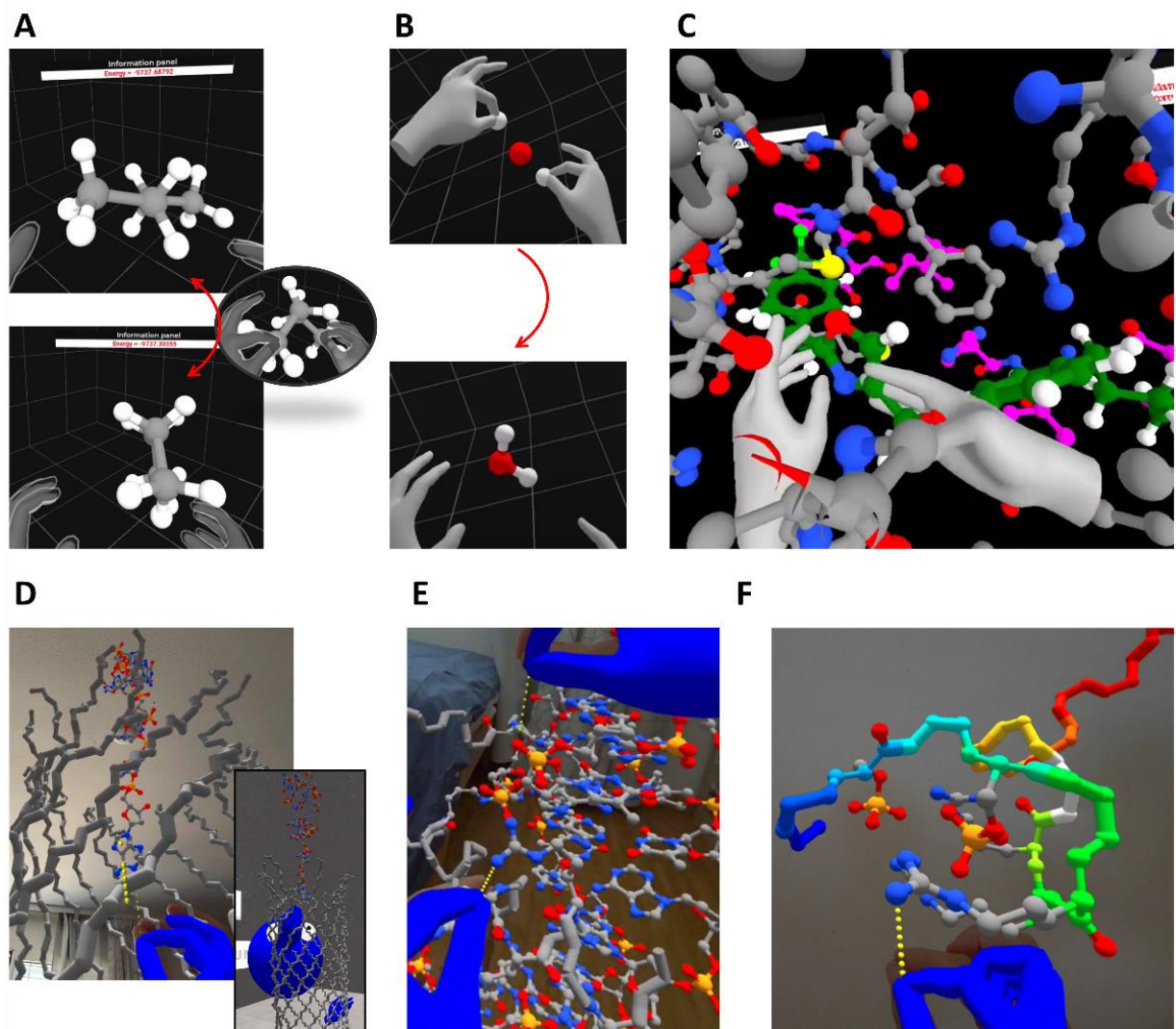

**Figure 7. Use cases of our HandMol programs.** (A) Driving the dihedral angle around the two central atoms of a butane molecule as ANI displays energies (red in the back). This is taken from a preset example of the HandMol prototype with which teachers can explain and students can explore by themselves various aspects of molecular dynamics and conformations, for example by comparing butane with 2-butene, isolated saturated rings with fused rings and aromatic rings, etc. (B) Building small molecules with ANI-2x and checking their geometries, to teach/learn about bonding, geometries, and more—as well as to build molecules that can then be exported for work outside of the app. In this specific case, the user approached two H atoms to an O atom in a linear arrangement; upon release ANI minimized the system building a water molecule with the right shape. (C) A small molecule (green carbons) right after it was docked into a pocket in a protein using the HandMol prototype. This was part of a quick exploratory work done for a grant proposal submission. (D) Threading a strand of DNA into a protein nanopore in an interactively steered MD simulation assisted by the Amber14 forcefield inside the modern HandMol, as seen from the headset (main panel) and from the side by a user at a computer (inset). The result of this operation was a system that was later on used as the starting point for MD simulations testing how much the DNA molecules blocks ion transport across the nanopore. (E) Probing docking strength at a protein-DNA complex, here pulling a positively charged arginine to break its interaction with a DNA phosphate group. To achieve this view, the system was asked to show only the DNA and arginines as balls and sticks, leaving the rest as backbone representations only. This was part of a study testing how strong the different kinds of interactions feel. (F) Probing the binding of phosphorylated serines to arginine sidechains in an intrinsically disordered peptide that presumably folds upon phosphorylation. This used custom parametrization for the phosphorylated serines, and was part of a discussion on how to interpret observations from various experiments.

## 6. The MolecularWeb universe: current standing, impact, and future outlook

This book chapter summarized how by developing the MolecularWeb ecosystem we materialize an effort to make immersive molecular science more accessible, intuitive, interactive, and collaborative. By consistently leveraging the power and ubiquity of standard web technologies (HTML, JavaScript, CSS, WebGL, WebXR, WebRTC) and by progressively integrating cutting-edge elements such as those based on AI, we have journeyed from relatively simple marker-based AR for largely static visualizations to highly dynamic, multiuser immersive environments that incorporate real-time physics and connections to sophisticated computational chemistry backends.

What next, then? While the MolecularWeb ecosystem has made significant strides, several exciting frontiers remain. The "dream tool" for molecular sciences is a seamlessly integrated, fully immersive, intuitively controlled, computationally powerful, and universally accessible collaborative environment.

A key area for advancement is in pushing the boundaries of natural human-computer interaction. This includes extending controller-free hand tracking to commodity devices by leveraging web-based machine learning libraries like MediaPipe, a concept already prototyped in moleculARweb's "Markerless AR" playground. Further enriching this interaction, the role of AI assistants can be expanded beyond simple command interpretation towards a more proactive partner, potentially even capable of generating and executing code within its code—a powerful but challenging goal that comes with significant security considerations—or perhaps "materializing itself" inside the XR sessions as one more user who then interacts with one naturally—as a virtual character in the Holodeck. The ultimate sensory goal, the integration of haptic feedback to let users "feel" molecular forces, remains on the horizon, awaiting the maturation of consumer-level hardware compatible with web standards, although as discussed earlier humans seem to be good at "extrapolating" haptic feedback from visual cues.[59,60]

Beyond the user interface, the scientific heart of the MolecularWeb ecosystem might also need to evolve from a tool primarily for exploring known structures and modifying some of its elements to one for genuine discovery. This certainly requires adding more features, as already possible in other immersive environments most notably UnityMol,[17] displaying more information about the molecular system like regular molecular graphics programs do, adding tools to mutate residues, change rotamers, add or remove atoms, etc.; and possibly broadening the scope of integrated physics and chemistry by adding a wider array of simulation engines such as the MARTINI force field for coarse-grained simulations of biological systems, reactive potentials for simulating bond formation and breaking, specialized engines for materials science, and even quick quantum mechanical methods for high-accuracy studies of small molecules. Such advanced simulations demand also advances in the visualization methods, moving beyond simple ball-and-stick representations to include cartoon representations of proteins, dynamic molecular surfaces, and crucially, the ability to render and interact with volumetric data like cryo-electron microscopy maps, which would in turn open a window into extending our tools directly into modern structural determination workflows.

Perhaps the most critical future directions involve closing the loop between our virtual environment, experimental data, and proven efficacy. A powerful evolution is to enable the real-time calculation of experimental observables directly from the manipulated models. Imagine interactively fitting a molecule into a density map and seeing a cross-correlation score update in real-time, or adjusting a protein's conformation and instantly seeing how its simulated SAXS profile compares to experimental data. This would transform HandMol into an invaluable tool for hypothesis testing and model refinement.

Finally, as we build these increasingly powerful tools, we have a responsibility to rigorously evaluate their impact. While anecdotal evidence and user feedback strongly indicate enhanced engagement, formal, large-scale pedagogical studies are essential to quantify improvements in learning outcomes and retention. Likewise, systematic evaluation is needed to determine if these immersive and intuitive interfaces truly lead to faster, more effective, and more creative research.

The "MolecularWeb" universe, with HandMol at its current frontier, demonstrates that many key components of a next-generation platform for molecular science are now feasible. By continuing to innovate at the intersection of web technologies, immersive interfaces, AI, and human-computer interaction, and by remaining keenly responsive to the needs of the scientific and educational communities, the web can indeed serve as the foundation for a future where exploring, understanding, and teaching the complexities of the molecular world is more intuitive, engaging, and collaborative than ever before.

## 7. Acknowledgements

institutions are acknowledged for their help providing content and ideas for our platforms, as well as for testing and applying our tools to various problems and in various situations.

Several individuals who helped with initial prototypes, server setup, and the contribution of open libraries on top of which we built these systems, are also heartly acknowledged. Likewise, we thank various teachers, students and researchers who provided valuable feedback after testing the tools.

Finally, we acknowledge the Swiss National Science Foundation, Hasler Stiftung (Bern), and Werner Siemens Stiftung for financial support through multiple grants specific for projects of the MolecularWeb universe.## 8. References

1. Goodsell, D. S. Looking at molecules—An essay on art and science. *ChemBioChem* **4**, 1293–1297 (2003).

2. Verma, V. V., Vimal, S., Mishra, M. K. & Sharma, V. K. A comprehensive review on structural insights through molecular visualization: tools, applications, and limitations. *J. Mol. Model.* **31**, 1–15 (2025).

3. Wu, H.-K. & Shah, P. Exploring visuospatial thinking in chemistry learning. *Sci. Educ.* **88**, 465–492 (2004).

4. Kozlíková, B. *et al.* Visualization of biomolecular structures: State of the art revisited. in *Computer Graphics Forum* vol. 36 178–204 (Wiley Online Library, 2017).

5. Gillet, A., Sanner, M., Stoffler, D. & Olson, A. Tangible Interfaces for Structural Molecular Biology. *Structure* **13**, 483–491 (2005).

6. O'Connor, M. *et al.* Sampling molecular conformations and dynamics in a multiuser virtual reality framework. *Sci. Adv.* **4**, eaat2731 (2018).

7. El Beheiry, M. *et al.* Virtual Reality: Beyond Visualization. *J. Mol. Biol.* **431**, 1315–1321 (2019).

8. Abriata, L. A. Elements for molecular graphics and modeling "in the Holodeck". *Methods in Molecular Biology* (2026).

9. O'donoghue, S. I. *et al.* Visualization of macromolecular structures. *Nat. Methods* **7**, S42–S55 (2010).